\input harvmac
\noblackbox

\lref\dfactory{J. Harvey, S. Kachru, G. Moore, and E. Silverstein,
unpublished}

\lref\raman{
R.~Sundrum,
``Effective field theory for a three-brane universe,''
Phys.\ Rev.\ D {\bf 59}, 085009 (1999)
[hep-ph/9805471].
}

\lref\boussopol{
R.~Bousso and J.~Polchinski,
``Quantization of four-form fluxes and 
dynamical neutralization of the  cosmological constant,''
JHEP {\bf 0006}, 006 (2000)
[hep-th/0004134].
}

\lref\gukov{
S.~Gukov, C.~Vafa and E.~Witten,
``CFT's from Calabi-Yau four-folds,''
Nucl.\ Phys.\ B {\bf 584}, 69 (2000)
[hep-th/9906070].
}

\lref\dinesil{
M.~Dine and E.~Silverstein,
``New M-theory backgrounds with frozen moduli,''
hep-th/9712166.
}

\lref\tunneling{
S.~Dimopoulos, S.~Kachru, N.~Kaloper, A.~E.~Lawrence and E.~Silverstein,
``Small numbers from tunneling between brane throats,''
hep-th/0104239;
S.~Dimopoulos, S.~Kachru, N.~Kaloper, A.~Lawrence and E.~Silverstein,
``Generating Small Numbers by Tunneling in Multi-Throat Compactifications,''
hep-th/0106128.
}
\lref\gkp{
S.~B.~Giddings, S.~Kachru and J.~Polchinski,
``Hierarchies from fluxes in string compactifications,''
hep-th/0105097.
}

\lref\dinesei{
M.~Dine and N.~Seiberg,
``Is The Superstring Weakly Coupled?,''
Phys.\ Lett.\ B {\bf 162}, 299 (1985).
}

\lref\selftune{
S.~Kachru, M.~Schulz and E.~Silverstein,
``Self-tuning flat domain walls in 5d gravity and string theory,''
Phys.\ Rev.\ D {\bf 62}, 045021 (2000)
[hep-th/0001206];
N.~Arkani-Hamed, S.~Dimopoulos, N.~Kaloper and R.~Sundrum,
``A small cosmological constant from a large extra dimension,''
Phys.\ Lett.\ B {\bf 480}, 193 (2000)
[hep-th/0001197].
}

\lref\pd{
T.~Damour and A.~M.~Polyakov,
``The String dilaton and a least coupling principle,''
Nucl.\ Phys.\ B {\bf 423}, 532 (1994)
[hep-th/9401069].
}

\lref\dps{
S.~P.~de Alwis, J.~Polchinski and R.~Schimmrigk,
``Heterotic Strings With Tree Level Cosmological Constant,''
Phys.\ Lett.\ B {\bf 218}, 449 (1989).
}

\lref\joebook{
J.~Polchinski,
``String theory. Vol. 1: An introduction to the bosonic string,''
{\it  Cambridge, UK: Univ. Pr. (1998) 402 p};
J.~Polchinski,
``String theory. Vol. 2: Superstring theory and beyond,''
{\it  Cambridge, UK: Univ. Pr. (1998) 531 p}.
}

\lref\chamseddine{
A.~H.~Chamseddine,
``A Study of noncritical strings in arbitrary dimensions,''
Nucl.\ Phys.\ B {\bf 368}, 98 (1992).
}

\lref\edds{
E.~Witten,
``Quantum gravity in de Sitter space,''
hep-th/0106109.
}

\lref\tom{
T.~Banks,
``Cosmological breaking of supersymmetry 
or little Lambda goes back to  the future. II,''
hep-th/0007146.
}

\lref\willy{
W.~Fischler, A.~Kashani-Poor, R.~McNees and S.~Paban,
``The acceleration of the universe, a challenge for string theory,''
hep-th/0104181.
}

\lref\lenny{
S.~Hellerman, N.~Kaloper and L.~Susskind,
``String theory and quintessence,''
JHEP {\bf 0106}, 003 (2001)
[hep-th/0104180].
}

\lref\andy{
A.~Strominger,
``The dS/CFT Correspondence,''
hep-th/0106113.
}

\lref\tomwilly{
T.~Banks and W.~Fischler,
``M-theory observables for cosmological space-times,''
hep-th/0102077.
}

\lref\mn{
J.~Maldacena and C.~Nunez,
``Supergravity description of field theories on curved manifolds and a no  go theorem,''
Int.\ J.\ Mod.\ Phys.\ A {\bf 16}, 822 (2001)
[hep-th/0007018].
}

\lref\herman{
H.~Verlinde,
``Holography and compactification,''
Nucl.\ Phys.\ B {\bf 580}, 264 (2000)
[hep-th/9906182].
}

\lref\fs{
W.~Fischler and L.~Susskind,
``Dilaton Tadpoles, String Condensates And Scale Invariance,''
Phys.\ Lett.\ B {\bf 171}, 383 (1986).
}

\lref\shamsandip{S. Kachru, S. Trivedi,..., work in progress.}
\lref\gkpII{S. Giddings, S. Kachru, and J. Polchinski, work
in progress}

\lref\edchir{
E.~Witten,
``Fermion Quantum Numbers In Kaluza-Klein Theory,''
PRINT-83-1056 (PRINCETON)
{\it  IN *APPELQUIST, T. (ED.) ET AL.: MODERN KALUZA-KLEIN THEORIES*, 
438-511. (IN *SHELTER ISLAND 1983, PROCEEDINGS, QUANTUM FIELD THEORY AND
THE FUNDAMENTAL PROBLEMS OF PHYSICS*, 227-277)}.
}

\lref\hw{
P.~Horava and E.~Witten,
``Heterotic and type I string dynamics from eleven dimensions,''
Nucl.\ Phys.\ B {\bf 460}, 506 (1996)
[hep-th/9510209].
}

\lref\corrprin{
L.~Susskind,
hep-th/9309145;
G.~T.~Horowitz and J.~Polchinski,
``A correspondence principle for black holes and strings,''
Phys.\ Rev.\ D {\bf 55}, 6189 (1997)
[hep-th/9612146].
}

\lref\kutsei{
D.~Kutasov and N.~Seiberg,
``Noncritical Superstrings,''
Phys.\ Lett.\ B {\bf 251}, 67 (1990).
}

\lref\otherconstr{
C.~M.~Hull,
``Timelike T-duality, de Sitter space, large N gauge 
theories and  topological field theory,''
JHEP {\bf 9807}, 021 (1998)
[hep-th/9806146];
V.~Balasubramanian, P.~Horava and D.~Minic,
``Deconstructing de Sitter,''
JHEP {\bf 0105}, 043 (2001)
[hep-th/0103171].
}

\lref\KachruED{
S.~Kachru, J.~Kumar and E.~Silverstein,
Class.\ Quant.\ Grav.\  {\bf 17}, 1139 (2000)
[hep-th/9907038].
}

\lref\MyersFV{
R.~C.~Myers,
``New Dimensions For Old Strings,''
Phys.\ Lett.\ B {\bf 199}, 371 (1987).
}

\lref\AntoniadisAA{
I.~Antoniadis, C.~Bachas, J.~Ellis and D.~V.~Nanopoulos,
``Cosmological String Theories And Discrete Inflation,''
Phys.\ Lett.\ B {\bf 211}, 393 (1988).
}

\lref\BanksZY{
T.~Banks,
``On isolated vacua and background independence,''
hep-th/0011255.
}

\lref\ArkaniKX{
N.~Arkani-Hamed, S.~Dimopoulos and J.~March-Russell,
``Stabilization of sub-millimeter dimensions: 
The new guise of the  hierarchy problem,''
Phys.\ Rev.\ D {\bf 63}, 064020 (2001)
[hep-th/9809124].
}

\lref\SilversteinKK{
E.~Silverstein,
``Gauge fields, scalars, warped geometry, and strings,''
Int.\ J.\ Mod.\ Phys.\ A {\bf 16}, 641 (2001)
[hep-th/0010144].
}

\lref\KachruTG{
S.~Kachru,
``Lectures on warped compactifications and stringy brane constructions,''
hep-th/0009247.
}

\lref\tommichael{T. Banks and M. Dine, work in progress.}


\def\frac#1#2{{#1 \over #2}}

\Title{\vbox{\baselineskip12pt\hbox{hep-th/0106209}
\hbox{SLAC-PUB-8869}
\hbox{NSF-ITP-01-}}}
{\vbox{ \centerline{(A)dS Backgrounds from Asymmetric Orientifolds}}}
\bigskip
\bigskip
\centerline{Eva Silverstein}
\bigskip
\centerline{{\it Department of Physics and SLAC, Stanford
University, Stanford, CA 94305/94309}} \centerline{{\it
Institute for Theoretical Physics, University of California, Santa
Barbara, CA 93106}} 

\bigskip
\noindent 

I present asymmetric orientifold models which, with 
the addition of RR
fluxes, fix all the NS NS moduli including the dilaton.
In critical string theory, this gives new AdS backgrounds with
(discretely tunably) weak string coupling.  
Extrapolating to super-critical
string theory, this construction 
leads to a promising candidate for a metastable
de Sitter background with string coupling of order 1/10 and dS radius
of order 100 times the string scale.  Extrapolating
further to larger and larger super-critical dimension suggests
the possibility of finding de Sitter backgrounds with 
weaker and weaker string coupling.  This note is an updated
version of the last part of my Strings 2001 talk.

\bigskip
\Date{June 2001}

\newsec{Introduction}

Because of bounds on Brans-Dicke forces and on time-dependence of
couplings, it is of interest to fix the moduli in string/M theory.
The diverse ingredients arising in modern string backgrounds, including
branes and RR fields, introduce new sources of moduli as well as new forces
which can help stabilize the moduli.   

In \S2\ I will present a six-dimensional model where the NS-NS moduli
(including the dilaton) are fixed, so that there are no
runaway directions in moduli space.  The strategy, as
outlined in the last few minutes of my Strings 2001 talk, 
is to balance the first few terms in string perturbation theory
off of each other by introducing large flux quanta and/or
brane charges, in such a way that a minimum arises
in the effective potential in a controlled regime at
weak string coupling.  The model of \S2\ has a minimum
of the dilaton potential below zero, producing an $AdS_6$
vacuum (with a string-scale compactification from ten
to six dimensions).  This sort of approach has been
studied in effective field theory in \raman\ArkaniKX\ and in general
terms in string theory in e.g. \gukov\boussopol; similar models
have been constructed geometrically in \gkp\gkpII\shamsandip.          

In \S3\ I will present a proposal for a construction of 
stringy dS space
which involves much less well understood {\it noncritical} string theory.
In particular, I will show how the naive scalar potential
of supercritical string theory on an asymmetric orbifold in
the presence of orientifold-antiorientifold pairs and RR fluxes
leads to a minimum of the scalar potential above zero.  In the
simplest model I consider, starting from dimension $D=12$, the
$dS_4$ radius that results is only larger than the string
length scale by two orders of magnitude, and
the string coupling $g_s$ is of order 1/10, so the construction depends on
the absence of coefficients at higher orders in perturbation
theory that might compensate these small but not arbitrarily
small suppression factors.  Modulo this issue, 
the closeness to the string scale
raises the interesting possibility that a cosmological analogue of
the correspondence principle \corrprin\ 
might account for the entropy suggested
by the area of the cosmological horizon.

We will also see that extrapolating this construction
to large super-critical dimensionality seems to give
better control, in that the string coupling can be
made weaker and weaker.  This might jive with
the idea that one should require more and more
degrees of freedom to describe a dS minimum as
it gets closer to a general relativistic regime \tom.  

However, it should be emphasized
that it is not completely known how to calculate in
non-critical dimensions, and I will make one or two assumptions 
along the way (which
I believe are plausible given the older results of
e.g. \dps\kutsei\chamseddine).  A complete worldsheet
description of the backgrounds here would require explicit
use of the Fischler-Susskind effect \fs\ since the 
dilaton is fixed by playing different orders of
perturbation theory off of each other.  Because
the candidate $dS$ minimum we find is only metastable,
it would probably be necessary to also understand the linear (in time)
dilaton theory to which it non-perturbatively decays,
which is the solution more traditionally studied, in order
to obtain a completely satisfactory construction.  However,
perturbatively it might be interesting to study the metastable
$dS$ minimum in its own right.          
  
In \dinesei, the problem of runaway directions toward weak coupling
limits of moduli space was identified, assuming no miraculous
small coefficients appear in the expansion around weak coupling.  
One way around that
is to minimize the moduli at strong coupling, as was argued for some examples
in \dinesil.  Another, as explored for example in the present
work and previous works such as 
\raman\ArkaniKX\gukov\gkp, is to introduce large ratios of coefficients
via discrete choices of flux quanta and brane and
orientifold charges.       

This latter freedom is perhaps unsettling because it expands
the arbitrariness of the vacuum (the vacuum degeneracy problem).
However, it has the redeeming feature that it expands the 
model-building possibilities enough
to plausibly provide quantum vacua with small cosmological constant,
as argued in e.g. \selftune\KachruTG\SilversteinKK\boussopol.  
As in these works, we will here not be
able to provide any rationale for why the initial conditions of
the universe favor the choices of discrete parameters that we use in
our constructions, and in fact our constructions will not
be realistic to begin with.  I feel it is nonetheless useful to obtain as
concrete a handle as possible on the range of possibilities,
particularly before declaring either victory or defeat.

In that vein, before proceeding to the constructions, I would like to 
make a few comments
on recent discussions in the literature on the issue of dS space
(or more generally accelerating universes).  
On the one hand, we have observational evidence for a currently
accelerating universe (with the dark energy just now having become
commensurate with matter).  On the other hand, if 
this were to persist indefinitely,
then the dark energy would dominate in the future, and one
would have a situation with cosmological event horizons with the
concommitant difficulties involved in formulating 
observables (in particular the lack of an S-matrix) 
\tom\edds\lenny\willy.\foot{Another argument often used against
the possibility of having dS space in M theory involves
the ``no go'' theorem of \mn.  However, the arguments
there were directed not just at dS space, but at nontrivial
warped compactifications including ones preserving
Poincare invariance.  In \mn, the assumptions
that went into this theorem are catalogued.
These include the absence of orientifold planes, whose
inclusion is known to yield warped compactifications
\herman\gkp. In our construction in \S3, both orientifold
planes and a tree-level potential coming from being away
from the critical dimension will be crucial, and these
are both elements of perturbative string theory
which are excluded by the assumptions of \mn.  
It seems to me that  ``no-go'' arguments based mostly on supergravity
(or on supergravity plus a partial list of known resolved
singularities)--while 
very useful for pointing toward the need for 
non-supergravity model-building ingredients--
are likely to be
unreliable as indicators of what physics is possible
within the full theory (as other examples in the
past have illustrated \edchir\hw).}  
These difficulties may well have resolutions
\tomwilly\andy\edds, but if they do not, one can still have perfect
agreement with observations in a universe which will tunnel
or roll out of a dS phase in the future.  In inflationary
scenarios, there are mechanisms for our universe to have 
exited from inflation in the past,
and it seems to me that
a similar mechanism could occur in the future to prevent
eternal acceleration.  

Because of the tendency of string-theoretic effective
potentials to decrease toward weak coupling \dinesei, 
in the dS construction we will
explore here, the minimum is indeed only metastable and would not
pose as great a problem in terms of observables.  However, as discussed
by T. Banks, if it holds up (i.e. if the perturbative
description does not break down),  it may provide an inroad
into the problem of understanding how a finite number 
out of the infinite available number of
degrees of freedom arrange themselves to describe the physics
associated with the dS minimum (though he points
out that the dS minimum may not
be accessible from the linear dilaton
regime after all for reasons similar to those discussed in 
\BanksZY\tommichael\ for other backgrounds).  Other proposals for
dS constructions have appeared in 
\otherconstr.\foot{Arguments against accelerating
cosmologies have also been
made based on the fact that no satisfactory dS background
has been built in string/M theory.
To me such arguments about what has (not) been accomplished
to date are weakened by 
the fact that there has been a preponderance of effort in the field directed 
(understandably) toward 
backgrounds with unbroken SUSY.  It is also hard to make precisely the 
(supersymmetric) Standard
Model in string theory (particularly including appealing
elements such as unification of couplings), 
and it hasn't quite been done, but few would
argue at this point that the absence of such a construction portends
a serious clash between string theory and experiment.  The problem
is messy, and because of the many moduli in string
theory the problem of building a model describing even  
a temporarily accelerating
universe is also grungy and may have resisted solution
for that reason alone.
In particular, one would expect the problem of fixing
moduli in systems with low-energy supersymmetry breaking
to require detailed knowledge of the full superpotential
and Kahler potential in the regime of the minimum.  These
quantities are still in the process of being computed in
most M-theoretic realizations of $4d$ N=1 supergravity.
}               
(This work is 
related to comments made in the last few minutes of my Strings
2001 talk; the material in the bulk of that talk can be found
in \tunneling.  There are independent
and probably more elegant geometrical constructions 
of $AdS$ and/or $dS$ backgrounds in progress 
by other authors \gkpII\shamsandip). 

\newsec{An AdS orientifold model in six dimensions}

I will here present a perturbative string model in which
all the non-periodic (i.e NS-NS) moduli are fixed, so that there are no
runaway directions.  A subset of this model was studied in \dfactory.    

Begin with type II string theory on a square $T^4$ at the self-dual
radius $R$.  Mod out by the asymmetric 
orientifold group generated by the following  
actions on the left and right movers of the string:
\eqn\genone{g_1: (0,s^2)_1(0,s^2)_2(0,s^2)_3(0,s^2)_4}
\eqn\genwo{g_2: \Omega I_4}
\eqn\genhree{g_3: (-1)^F (0,s^2)_1(0,s^2)_2}
\eqn\genfour{g_4:  (-1)^{F_R}(-1,s^2)_1(-1, s^2)_2(-1,1)_3
(-1,1)_4}

Here $s$ represents an asymmetric shift so that $(0,s^2)_j$ acts
as $(-1)^{m_j+n_j}$ where $m_j$ and $n_j$ are momentum and winding numbers
on the $j$th circle.  $I_4$ denotes reflection on four coordinates.  
The first of these actions \genone\ generates the $SO(8)$ lattice from
the $SU(2)^4$ lattice we started with, which is needed
here for level-matching.  The elements $g_2$
and $g_2g_3$ introduce orientifold 5-planes and anti-orientifold
5-planes at locations separated by the shifts.  Important combinations
of these elements include space-filling orientifolds and antiorientifolds
$\Omega (-1)^F, \Omega (0,s^2)^2$ and an element
\eqn\Kthree{h_1=I_4(-1)^F.}
Let us choose the discrete torsion such that 
$g_3$ projects out the NS-NS scalars from the $|g_3g_4>$
and $|g_4>$ twisted sectors (these are compatible in
this model).  
Let us also choose the discrete torsion such that $h_1$ kills
the twisted gravitini from the $|g_3g_4>$ twisted sector,
and such that $g_3g_4$ projects out the NS-NS scalars
from the $|h_1>$ twisted sector (these are compatible
in this model, with the $2\pi$ rotations implicit
in the $(-1)^F$ actions acting in internal compactified directions,
a distinction that affects the twisted sectors).

With these specifications, this model has no NS NS moduli.  
The element $g_4$ projects out all the untwisted NS-NS moduli of
the torus.  None
reemerge from twisted sectors.  
In the $g_3$ sector, the lightest
states are massless fermions.  In the $g_4$ sector, one finds
NS-NS moduli that are killed by the above choice of discrete torsion,
as is the case in the $|h_1>$ sector.  Similarly, in other
sectors obtained by products of group elements (which
are isomorphic to those already discussed), potential NS-NS moduli
are projected out when they arise.    

This theory includes RR fields.  RR moduli do not lead to
runaway behavior since they are periodic, and they
only couple derivatively to the rest of the theory.  The RR field
strengths will also be important in the construction.  
In particular, we can include combinations of RR fluxes
which respect the orbifold symmetries.

Including these ingredients, 
the potential energy of this theory is, in string frame
\eqn\Vstr{
V_s(g_s) \sim -32 {T_O\over g_s} + F_{RR}^2 + {\cal O}(g_s)
}
where $F_{RR}$ encodes the RR fluxes (each type of which
integrates to an integer $Q_{RR}$ over cycles in
the compactification which survive the orientifolding) and $T_O/g_s$ is
the tension of a single orientifold or antiorientifold
5-plane (this first term in \Vstr\ also then includes the 
contribution of the T-dual
$O9$-planes and $\bar O9$-planes).  
We are here taking the flux quanta to be large
so that the one-loop contribution to the vacuum energy
is dwarfed by the $F_{RR}^2$ contribution to the potential
energy at order $g_s^0$.  
In the six-dimensional Einstein frame, the potential is
\eqn\VEin{
V_E(g_s)\sim -32 {T_O}g_s^2 + F_{RR}^2g_s^3+{\cal O}(g_s^4)
}    

From \VEin\ we see that if we introduce a large number of RR
flux quanta, we can balance the first and second terms in this
perturbation expansion off of each other and obtain a stable
minimum of the potential energy \raman\boussopol.  This minimum has negative
potential energy and leads to an $AdS_6$ spacetime in
the noncompact dimensions.  It would be interesting to understand the dual
$5d$ 
CFT determined by the string theory on this background.  Since
here the internal space is string-scale, there is no large
``sphere'' or more general Einsteins space 
component of the geometry, so
the matter content and symmetries of this case are quite
different from other examples of AdS backrounds.     

Because here $1 << 1/g_s \sim Q_{RR}^2 << 1/g_s^2$, 
the curvature AdS space
in this model is much smaller than string-scale curvature
$m_s^2$ (and therefore also much smaller than
the compactification scale in this model).  The mass of
the dilaton is also much smaller than $m_s$.  
In realistic applications, this mass must be at least as
big as an inverse millimeter (or nearly decouple at long
distance for some other reason \pd).  Of course this model
is not realistic in any case because, because of the dimensionality
and the sign of the cosmological constant (among other things).

\newsec{A ``noncritical'' approach to dS space}

Since all observational indications are that we have, at
least currently, a positive
cosmological term (scalar potential) with the dark energy having
an equation of state $p/\rho \equiv w \le -0.66$,
it is of interest to either 
locally fix the moduli at a point where the potential
is positive or find an extroardinarily flat potential
to describe ``quintessence''.  
One typically finds runaway behavior toward weakly-coupled
boundaries of moduli space \dinesei, so that the simplest possibility
is a local minimum above zero obtained by playing the
first three terms in perturbation theory off of one another.  
In this section we will discuss such a minimum in the
context of noncritical string theory.  

I will take the point of view (see e.g. \dps\kutsei\chamseddine)
that non-critical string theory may be formulated
in any dimension as long as one solves the 
resulting string equations
of motion in a reliable regime with a background that
has no perturbative instabilities.  In the simplest construction here, we will
have a perturbative description controlled by dimensionless
parameters of order 1/10 or 1/100, not ones that can be
made arbitrarily small for a fixed dimensionality, 
so I will not be able to prove
the background exists.  I will simply assume that 
these ratios are sufficient to 
give a reliable perturbation expansion.   
However, even
given this, it is worth saying from the start
that the resulting background is not realistic due
to the scales that emerge.  Still, it is potentially
worth pursuing this sort of solution because of the conceptual
issues involved in formulating and studying accelerating
cosmologies.   

Consider a background with dimension $D$ not necessarily
equal to the critical dimension $D_c=10$ and with orientifold
planes and RR field strengths.    
In D-dimensional string frame, for regimes where effective
field theory applies, one has an action \dps\joebook\  
\eqn\strfourd{\eqalign{
& S_{string}={1\over {2\kappa_0^2}}\int d^Dx\sqrt{-G_s}e^{-2\Phi}
\biggl[ R-{{2(D-D_c)}\over{3\alpha^\prime}}\biggr]\cr
& -\sum_{O_p} T_{O_p}\int d^{p+1}\eta e^{-\Phi}\sqrt{-\gamma_s}\cr
& - {1\over{4\kappa_0^2}}\int d^Dx\sqrt{-G_s}
\sum_{F_{RR}}|F_{RR}|^2\cr
}}
where the second line involves a sum over all orientifold planes
that are present in the background and the third line involves
a sum over all RR field strengths $F_{RR}$ that are turned on
in the background.  Here so far the D-dimensional metric $G_{s}$ and
the brane metric $\gamma_s$ are in the string frame. 

In the absence of the last two lines of \strfourd, one finds
a linear dilaton background to solve the dilaton equation
of motion arising from \strfourd, a solution which is
in fact exact in $\alpha^\prime$ \MyersFV\AntoniadisAA.  
Much of the literature
on noncritical string theory involves this solution.  
However in the presence of the other terms, there is
a priori the possibility of other solutions.  In particular,
here we will argue that there are compactified solutions
with meta-stabilized dilaton.  


If we dimensionally reduce to four dimensions and switch to
Einstein frame, we obtain an effective potential for $\Phi$
of the form 
\eqn\potI{
U(\Phi)\propto \biggl(ae^{2\Phi}-b e^{3\Phi} + ce^{4\Phi}\biggr)
}
where $a$, $b$, and $c$ are positive constants read off from the first three
lines of \strfourd\ which will be specified explicitly below given $T_{O_p}$.  
Extremizing this gives
\eqn\extrema{
e^{\Phi_\pm}={{3b\pm\sqrt{9b^2-32ac}}\over{8c}}
}    
If this potential is reliable, and if 
\eqn\condI{
9b^2>32ac
}
then there is a metastable minimum of the dilaton potential at $\Phi=\Phi_+$.

The other moduli of the string background can be stabilized in 
a manner similar to that employed in \S2.  In fact, starting
in $D=12$ dimensions we can consider a simple generalization
of the model of
\genone - \genfour.  That orbifold/orientifold group acted on
four dimensions, giving a reduction of $10d$ critical
string theory down to $6d$.  With the same action on twice as many
coordinates, we can reduce from $D=12$ to $D=4$, and again fix
all the NS-NS moduli, leaving invariant some combinations of
Ramond fields.  This leaves us with a string-scale compactification
manifold, and we can put Ramond flux along diagonal directions
of the torus so as to yield a small separation of scales between
the RR $q$-form field strengths and the inverse radii of the corresponding
cycles.  This gives 
\eqn\cdef{
c\sim {1\over 4}\sum{{(Q_{RR}^{(q)})^2}\over V_q}
}
where $V_q$ is the volume in string units
of the cycle carrying $q$-form flux $Q_{RR}^{(q)}$.

In order to determine $b$ in \potI, we need to know the
orientifold tensions $T_{O_p}$ in \strfourd.  These will in general
be a function of $D$ (or equivalently $d\equiv D-2$) and
the spatial dimension $p$ of the $O_p$-plane.  We can start to
analyze this by generalizing the calculations of D-brane tensions     
\joebook\ to arbitrary dimension, following a similar
analysis \chamseddine\ of the closed string partition
function and spectrum in noncritical dimensions.  
In the case of the superstring,
this gives an NS-NS exchange between Dp-branes separated
by a distance $y$
\eqn\NSexch{
{\cal A}_{NS-NS}=i{{V_{p+1}}\over{(8\pi^2\alpha^\prime)^{{p+1}\over 2}}}
\int_0^\infty dt t^{-{{(p-d+3)}\over 2}} e^{-{{ty^2}
\over{2\pi\alpha^\prime}}}
e^{{{d\pi}\over {8t}}}(d e^{-\pi/t}+\dots)
}
where $\dots$ represents an infinite series of terms suppressed
relative to the first term
in the $t\to 0$ infrared limit in the closed-string channel.  
For the critical superstring case of $d=8$, this reduces to
the formula (13.3.1) in \joebook, in which this
leading exchange is massless.  More generally, it describes 
noncritical strings propagating on a linear dilaton background,
as in the corresponding closed-string analysis of partition
functions in \chamseddine.  In sub-critical dimensions
$d<8$, the mode is effectively massive due to the linear
dilaton, corresponding to the surviving exponential
suppression $e^{{\pi\over t}{(d-8)\over 8}}$.  On the other
hand, for $d>8$ one finds the leading modes to be
effectively tachyonic in the linear dilaton background.        
However, as discussed in \chamseddine, in an effective
action description the fields all have nontachyonic 
$m^2$ as in the critical string; the instability 
for $d>8$ arises
as a result of the effect of the dilaton coupling on the
equation of motion.  

We are interested in looking for solutions 
\extrema\ in which the dilaton is fixed.  In such
a background, there is no linear dilaton and
the masses are not even effectively tachyonic.  
I will now assume that in fact
the only effect of stabilizing the
dilaton on the calculation of D-brane and orientifold tensions
is to shift the effective
masses of the exchanged particles in \NSexch,
in particular shifting the graviton exchange to the
massless level.  In particular we will assume
it does not change the normalization of the
amplitude.  In particular, this amounts to assuming
that the 
multiplicities of particles, particularly the low-lying
ones like 
the graviton, do not change when one considers
a different solution of the dilaton equation of motion,
which appears to me a reasonable assumption given that
the effective action appears self-consistently reliable in
the models we will study.    

Doing this and comparing the result to the appropriate
low-energy Greens function in D dimensions gives the result
\eqn\Pten{
T_p={{2^{4-d/4}\pi^{1/2}}\over{\kappa_0(2\pi{\alpha^\prime}^{1/2})^{p+1-d/2}}}
}
for the Dp-brane tension.  One finds for the orientifold tensions
\eqn\DOreln{
2^{(d+2)-(p+1)}T_{Op}=-{{2^{4+d/4}\pi^{1/2}}
\over{\kappa_0(2\pi{\alpha^\prime}^{1/2})^{p+1-d/2}}}
}
where we have included all $2^{(d+2)-(p+1)}$ orientifold p-planes present
in $d+2$ dimensions on a torus.

In our model with $D=d+2=12$, 
let us take $\kappa_0=(4\pi^2\alpha^\prime)^{5/2}$.  
$2\pi(\alpha^\prime)^{1/2}=2\pi R_{sd}$ is the linear size of the
self-dual compactification manifold, so this amounts to setting the
dimensionful coupling $\kappa_0$ to be at the self-dual 
compactification scale.  
The choice of $\kappa_0$ does not affect any physics.  However,
rescaling $\kappa_0$ does rescale the coefficients in the
string coupling expansion, and in our problem
we will not have the luxury of parametrically lowering the dilaton
to achieve an arbitrarily well controlled series.
Instead, as we will see, 
with this choice of $\kappa_0$, our string coupling will be of order 1/10.  
Because we picked a natural scale for $\kappa_0$, I expect generically
the coefficients in the $g_s$ expansion to be of order one, so that
our use of the first three terms only in will be valid.  However,
the possibility remains that large coefficients arise at
higher orders in perturbation theory in which case our approximation
would fail.  In addition to factors of $g_s\sim 1/10$, higher
orders in the loop expansion come with powers of $1/2\pi$ from
loop momentum integrals, which aids our cause.\foot{On the other hand,
another natural choice for $\kappa_0$ might be to take
$\kappa_0=R_{sd}^5={\alpha^\prime}^{5/2}$.  Doing this would
cancel the powers of $1/2\pi$ from the loop momentum integrals,
leaving us still with the suppression by powers of 1/10 only.}  
In any case however, because of
the fact that there are no arbitrarily negative contributions
in the $b$ term at fixed $d$, we cannot tune the dilaton to be arbitrarily
weak, and so cannot be completely sure that the candidate minimum
we will study exists.  

It is interesting to contemplate extrapolating
to large dimensionality, in which case the negative contribution
from \DOreln\ grows exponentially whereas the leading
positive term grows only linearly \strfourd.  This may lead to
better control, though going very far from the critical dimension
might introduce even more subtleties.      

The above results and assumptions 
lead to the following contributions to the effective
potential in four-dimensional Einstein frame:
\eqn\fourdE{
S_E=\int d^4 x\sqrt{g_E}{1\over{(4\pi^2\alpha^\prime)^2}}
\biggl[ae^{2\Phi}-be^{3\Phi}+ce^{4\Phi}\biggr]
}
where $a=8\pi^2/3$, $b=\pi^{1/2}2^{15/2}$ (including a
factor of two for the fact that there are two sets of
orientifolds in our model) and $c$ is given by \cdef.  

Let us choose the RR flux to get the smallest possible value of the
string coupling $e^{\Phi_+}$.  This means taking $c$ such
that $32ac\sim 9b^2$, so that 
\eqn\dilfin{
e^{\Phi_+}\sim {{4a}\over{3b}}\sim 0.11.  
}
Given this, we find the cosmological constant at the
minimum, $\Lambda\equiv U(\Phi_+)$, to be bounded by
\eqn\cc{
\Lambda \sim {1\over{(2\pi R_{sd})^4}}(0.05)
}
so that the curvature radius $L$ of the dS space is greater than
string scale $(\alpha^\prime)^{1/2}$ 
by two orders of magnitude (using the
relation that the general relativistic
dimension two cosmological constant is given
by $\Lambda_{GR}=1/L^2=\Lambda/M_4^2
=(1/\alpha^\prime)(0.03/4\pi^2)$, reading
off $M_4$ from the $4d$ Einstein frame action).  

Clearly, establishing completely the validity (and utility)
of this approach to constructing dS space will require getting
a handle on the worldsheet description of the CFT including
the Fischler-Susskind \fs\ description of the
competition between different orders in perturbation
theory that is central to the mechanism.  If this can
be accomplished, it is intriguing that increasing
the dimensionality (and therefore the naive
number of degrees of freedom) seems to give a better and better
perturbative description of dS space.

\noindent{\bf Acknowledgements}

I would like to thank the organizers of Strings 2001 for
organizing such a stimulating conference.
I would like to thank T. Banks, R. Bousso, S. Giddings, 
S. Gukov, W. Fischler, S. Hellerman, G. Horowitz, 
S. Kachru, N. Kaloper, J. Maldacena,
J. Polchinski, A. Strominger, L. Susskind, and E. Witten, 
for interesting discussions on this and/or related
topics.  The
upcoming work \gkpII\ has
some overlap in technique with our work here at least
with regard to AdS vacua, and
I thank those authors for discussions.  
I would like to thank the Institute for Theoretical Physics
at UCSB for hospitality and support, 
and the DOE (contract DE-AC03-76SF00515 and OJI) and
the Sloan Foundation for support.

\listrefs

\end